\documentclass{appolb}
\usepackage{graphicx}
\usepackage{cite}

\begin{document}
\title{PHENIX results on Bose-Einstein correlation functions%
\thanks{Presented at the XI Workshop on Particle Correlations and Femtoscopy, \\3-7 November 2015, Centre for Innovation and Technology Transfer Management, Warsaw University of Technology
}
}
\author{D\'aniel Kincses for the PHENIX Collaboration
\thanks{This research was supported by the Hungarian OTKA NK101438 grant, as well as by the funding agencies listed in ref. \cite{funding}.}
\address{E\"otv\"os Lor\'and University, Hungary}
}
\pagestyle{plain}
\maketitle
\begin{abstract}

Measurement of Bose-Einstein or HBT correlations of identified charged particles provide insight into the space-time structure of particle emitting sources in heavy-ion collisions. In this paper we present the latest results from the RHIC PHENIX experiment on such measurements.

\end{abstract}
\PACS{25.75.Dw}

\section{Introduction}

The PHENIX experiment at the BNL Relativistic Heavy Ion Collider (RHIC) has collected comprehensive data in multiple different collision systems from p+p, p+Au, d+Au, He+Au through Cu+Cu to Au+Au and U+U collisions, at energies that are varied in the region where the transition from hadronic to quark matter is expected to occur (7.7 GeV to 510 GeV). The importance of this beam energy scan program is that comparing results at these different energies allows us to investigate the structure of QCD matter, and the quark-hadron transition. One of the best tools to gain information about the particle-emitting source is the measurement of Bose-Einstein or HBT correlations, and in this paper we present the latest PHENIX results of such measurements.

\newpage
\section{Comparison of charged pion and kaon femtoscopy}

\begin{figure}[htb]
\centerline{%
\includegraphics[width=12.cm]{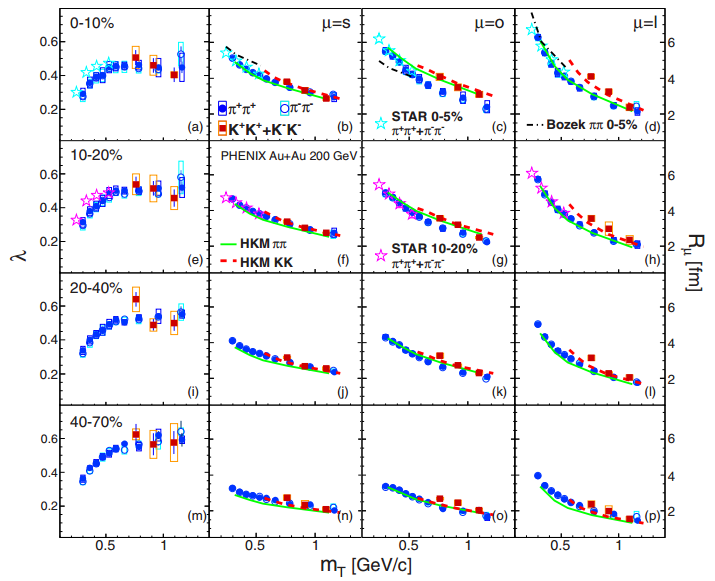}}
\caption{Extracted HBT parameters of charged pions and kaons as a function of $m_T$ for the centralities indicated. Results of charged pions from STAR \cite{Adams:2004yc} are compared. Calculations from the hydrokinetic model (HKM) \cite{Karpenko:2012yf} and viscous-hydrodynamic model (Bozek) \cite{PhysRevC.89.044904} are also shown.\newline}
\label{Fig1}
\end{figure}

Figure \ref{Fig1}. shows the azimuthal-integrated Gaussian HBT parameters of charged pions and kaons 
in Au+Au collisions at $\sqrt{s_{NN}} = 200$ GeV, for four centrality classes as a function of $m_T$  as measured recently by PHENIX \cite{Adare:2015bcj}. Results for charged pions in the low $m_T$ region from STAR \cite{Adams:2004yc} are also plotted. The source parameters from the two experiments are in good agreement, but the PHENIX $\lambda$ parameters are 20\% lower at low $m_T$ . A possible reason can be that the value of $\lambda$ is sensitive to the combinatorial background level, which may differ between PHENIX and STAR. Positive and negative pions are quite consistent. The presented data are also consistent with earlier PHENIX results \cite{Adler:2004rq, Afanasiev:2009ii}. The hydrokinetic model reproduces most aspects of the data of both charged pions and kaons, but it fails to accurately describe the difference in $R_o$.

\newpage
\section{Beam energy and system size dependence of HBT radii}

\begin{figure}[htb]
\centerline{%
\includegraphics[width=7.35cm]{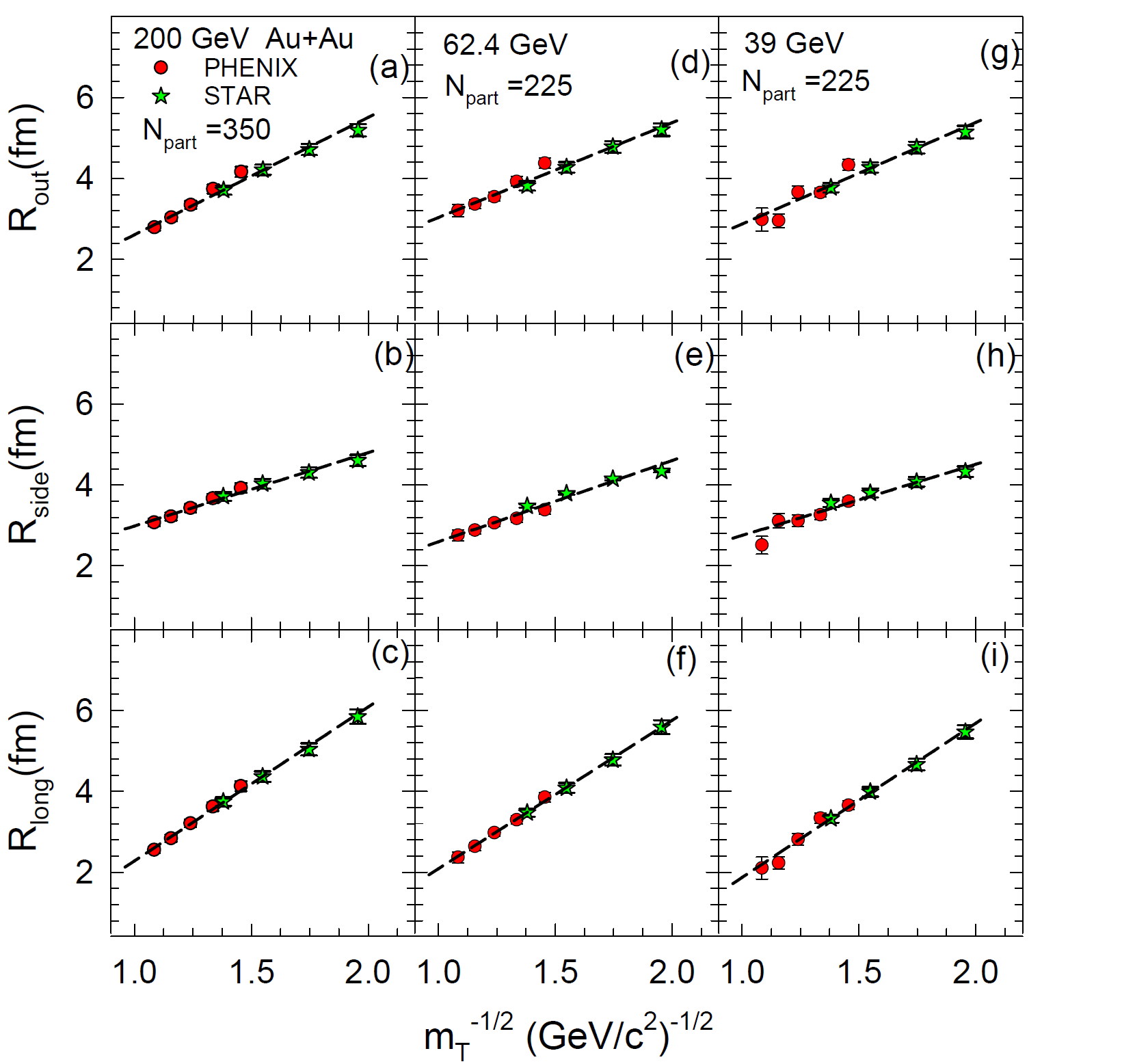}}
\caption{Comparison of PHENIX \cite{Adare:2014qvs} and STAR \cite{Adamczyk:2014mxp} HBT radii for Au+Au collisions at $\sqrt{s_{NN}}$ = 39.0, 62.4 and 200 GeV as indicated.\newline}
\label{Fig2}
\centerline{%
\includegraphics[width=6.27cm]{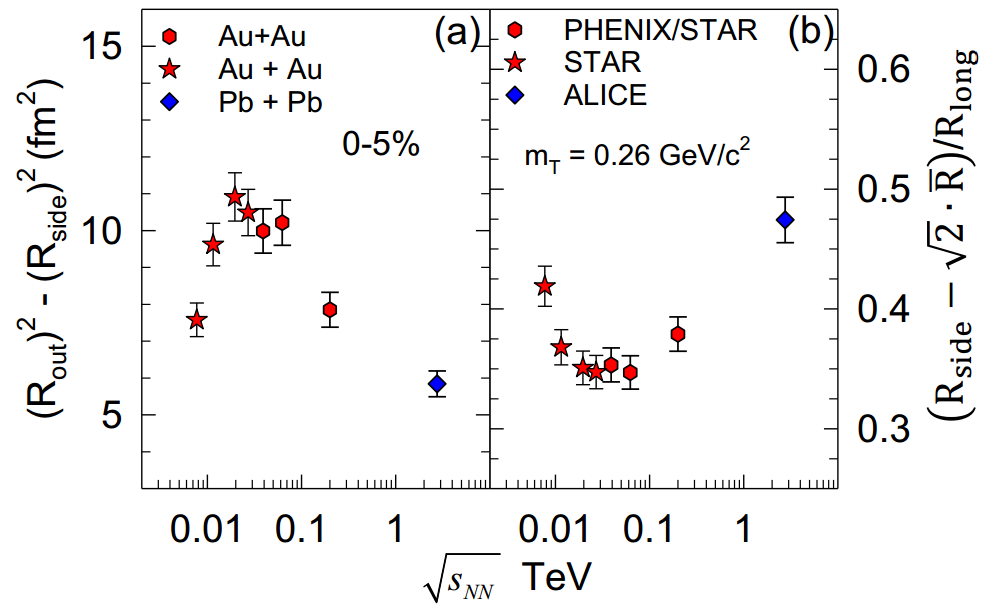}}
\caption{The $\sqrt{s_{NN}}$ dependencies of combinations of HBT radii sensitive to expansion velocity  (a) and emission duration (b) from \cite{Adare:2014qvs,Adamczyk:2014mxp,Kisiel:2011}.}
\label{Fig3}
\end{figure}

PHENIX also measured the Gaussian HBT radii in Cu+Cu and Au+Au collisions at several beam energies \cite{Adare:2014qvs}. The extracted radii, which were compared to recent STAR \cite{Adamczyk:2014mxp} and ALICE \cite{Kisiel:2011} data, show characteristic scaling patterns as a function of the initial transverse size and the transverse mass of the emitted pion pairs (see Fig.\ref{Fig2}.), consistent with hydrodynamic like expansion. On Fig.\ref{Fig3}. we can see specific combinations of the three-dimensional radii \cite{Adare:2014qvs,Adamczyk:2014mxp,Kisiel:2011} that are sensitive to the medium expansion velocity and lifetime, and the pion emission duration. These show non-monotonic $\sqrt{s_{NN}}$ dependencies, which may be an indication of the critical endpoint in the phase diagram of hot and dense nuclear matter.

\section{Ongoing work: PHENIX Levy HBT analysis and future plans}

In the previous section we have seen a measurement in which there may be an indication of the critical endpoint, but there may be an other way to find the CEP - the measurement of Levy exponent $\alpha_{Levy}$ at different beam energies. An ongoing PHENIX analysis is currently trying to make a detailed shape analysis of one-dimensional two-pion correlation functions with using Levy source instead of Gaussian on $\sqrt{s_{NN}}$=200 GeV Au+Au data. The Levy exponent $\alpha_{Levy}$ is actually identical to critical exponent $\eta$ which has a particular value $\eta=0.5$ at the critical point \cite{Csorgo:2003uv, Csorgo:2005it, Csorgo:2009gb}. Preliminary results from \cite{Csanad:2005nr} suggest that $\alpha_{Levy}$ at 200 GeV is above 0.5 but also less than 2, the traditionally assumed value in a Gaussian approximation. This ongoing analysis indicates values similar to that earlier PHENIX preliminary results and the plan is to repeat this measurement at lower energies - if $\alpha_{Levy}$ decreases, and reach the 0.5 value somewhere, it may also be an indication of the critical endpoint. The $m_t$ dependence of the $\lambda$ parameter (obtained from various fits) is also investigated in this ongoing analysis. A comparison will be done to previous PHENIX measurements \cite{Adare:2015bcj,Adler:2004rq,Afanasiev:2009ii,Adare:2014qvs,Csanad:2005nr} where Gaussian fit results were obtained (corresponding to the special case of $\alpha = 2$). Regarding $U_A(1)$ symmetry restoration \cite{Csorgo:2009pa}, the PHENIX preliminary analysis \cite{Csanad:2005nr} yielded consistent results with those already published \cite{Adare:2015bcj,Adler:2004rq,Afanasiev:2009ii,Adare:2014qvs}. The final analysis is expected to show a significant improvement in the statistical and systematic uncertainties.

\end{document}